\documentclass[11pt]{article}
\usepackage{a4wide}
\usepackage{amsmath,amssymb,amsfonts}
\usepackage{graphicx,subfigure,epsfig}
\usepackage{xcolor}
\usepackage[T1]{fontenc}
\usepackage{upquote,multirow,slashed}
\usepackage{cite,appendix}
\usepackage{empheq}
\usepackage{lipsum}

\definecolor{ForestGreen}{rgb}{0.15,0.70,0.15}
\usepackage[colorlinks=true]{hyperref}
\hypersetup{
 linktocpage=true,
 citecolor=ForestGreen,
 linkcolor=blue,
 urlcolor=blue}
\date{\vspace{-5ex}}

\numberwithin{equation}{section}

\newcommand{\ttbar}{t{\bar t}}
\newcommand{\bbbar}{b{\bar b}}

\newcommand{\as}{\alpha_s}

\newcommand{\afb}{A_{\rm FB}}

\newcommand{\sia}{\sigma_A}
\newcommand{\sis}{\sigma_S}

\newcommand{\mycomment}[1]{ }
\begin{document}

\title{\vspace{-2cm}
\vspace{2cm}
\textbf{Forward-backward asymmetries of the heavy quark pair production in $e^+e^-$ collisions at $\mathcal{O}(\alpha_s^2)$}}

\author{Long Chen\footnote{\textit{E-mail}: longchen@physik.rwth-aachen.de} \\ [.5cm]
Institut f\"ur Theoretische Teilchenphysik und Kosmologie, \\
    RWTH Aachen University,  52056 Aachen, Germany~\\
}

\maketitle

\begin{center}
{\small \textit{Talk presented at the International Workshop on Future Linear Colliders (LCWS2021)}} 
\end{center}
\begin{center}
{\small \textit{ 15-18 March 2021. C21-03-15.1}}
\end{center}

\noindent\rule{\textwidth}{.5pt}
\begin{abstract}
The talk is on a computational set-up for calculating the production of a massive quark-antiquark pair in electron positron collisions to order $\alpha_s^2$ in the coupling of quantum chromodynamics (QCD) at the differential level using the antenna subtraction method. 
Theoretical predictions on the production of top quark pairs in the continuum, and the bottom quark pairs at the $Z$ resonance, will be discussed. 
In particular, we would be focusing on the order $\alpha_s^2$ QCD corrections to the heavy quark forward-backward asymmetry (AFB) in electron positron collisions.
In the case of the AFB of bottom quarks at the $Z$ resonance, the QCD corrections are determined with respect to both the bottom quark axis and the thrust axis.
We will also briefly discuss improvements on these QCD corrections brought by applying the optimization procedure based on the Principle of Maximum Conformality.
\end{abstract}
\noindent\rule{\textwidth}{.5pt}

\thispagestyle{empty}



\thispagestyle{empty}
\vspace{1cm}

\allowdisplaybreaks


\section{Introduction}
\label{SEC:introduction}

The exploration of the physics of heavy quarks, that is, bottom and top quarks, is among the core physics issues at both the past and the future linear or circular electron-positron colliders~\cite{ALEPH:2005ab,Alcaraz:2009jr,AguilarSaavedra:2001rg,Baer:2013cma,Gomez-Ceballos:2013zzn}, such as the International Linear Collider (ILC)~\cite{Baer:2013cma}, the Compact Linear Collider (CLIC)~\cite{Linssen:2012hp}, the Future Circular electron-positron Collider (FCC-ee)~\cite{Abada:2019zxq} and the Circular Electron Positron Collider (CEPC)~\cite{CEPC-SPPCStudyGroup:2015csa,CEPC-SPPCStudyGroup:2015esa}.
Among the set precision observables associated with the heavy quark production at these lepton colliders include the forward-backward asymmetries, the key observables for the determination of the neutral current couplings of leptons and quarks in the reactions $e^+ e^- \to Q {\bar Q}$. 
As far as quarks are concerned, the most precisely known asymmetry is that of the $b$ quark at the $Z$ resonance $A_{\rm FB}^b$, which was measured at SLAC and LEP with an accuracy of 1.7 percent~\cite{ALEPH:2005ab,Alcaraz:2009jr}. 
The measured $A_{\rm FB}^b$ at the $Z$ resonance shows a relatively large deviation, about 2.9~$\sigma$, from the respective Standard Model (SM) fit~\footnote{It is noted that the deviation of this experimental measurement from the SM fit result given in a later ref.~\cite{Baak:2014ora} reduces to 2.5~$\sigma$, and to 2.4~$\sigma$ compared to the updated SM prediction in ref.~\cite{Tanabashi:2018oca}.}. 
So far, it has not been clarified whether this deviation is due to underestimated experimental and/or theoretical uncertainties or whether it is a hint of new physics.

At a future linear or circular $e^+ e^-$ collider, precision determinations of electroweak parameters will again involve forward-backward asymmetries. If such a collider will be operated at the $Z$ resonance, an accuracy of about 0.1 percent may be reached for these observables~\cite{Hawkings:1999ac,Erler:2000jg,ILD:2020qve}. 
At the International Linear Collider (ILC), the top-quark forward-backward asymmetry can be measured to a precision of below one percent in relative~\cite{Devetak:2010na,Amjad:2013tlv,ILD:2020qve}. 
Needless to say, precise predictions are required, too, on the theoretical side.

As far as the production of $\ttbar$, or more general, the production of a heavy quark-antiquark pair $(Q{\bar Q})$ in $e^+ e^-$ collisions in the continuum is concerned, differential predictions at next-to-leading order (NLO) QCD (corresponding to $\mathcal{O}(\alpha_s)$ for this process) have been known for a long time for $Q{\bar Q}$ \cite{Jersak:1981sp} and $Q{\bar Q}$ + jet~\cite{Bernreuther:1997jn,Brandenburg:1997pu,Rodrigo:1997gy,Rodrigo:1999qg,Nason:1997tz,Nason:1997nw} final states. 
Also the NLO electroweak corrections are known~\cite{Beenakker:1991ca,Bohm:1989pb,Bardin:1999yd,Freitas:2004mn,Fleischer:2003kk,Hahn:2003ab,Khiem:2012bp}. 
Recently, the QED initial state corrections to the forward-backward asymmetry for $e^+ e^- \to \gamma^*/Z^*$ are calculated in the leading logarithmic approximation to high orders in ref.~\cite{Blumlein:2021jdl}.
The total $Q{\bar Q}$ cross section $\sigma_{Q{\bar Q}}$ was computed to order $\alpha_s^2$ and order $\alpha_s^3$ in \cite{Gorishnii:1986pz,Chetyrkin:1996cf,Chetyrkin:1997qi,Chetyrkin:1997pn} and \cite{Kiyo:2009gb}, respectively, using approximations as far as the dependence of $\sigma_{Q{\bar Q}}$ on the mass of $Q$ is concerned.
A computation of the cross section and of differential distributions for $\ttbar$ production at order $\alpha_s^2$ with full top-mass dependence was reported in \cite{Gao:2014nva,Gao:2014eea} using a hybrid approach combining phase-space slicing and the dipole subtraction, and also in~\cite{Chen:2016zbz} using the antenna subtraction method~\cite{Kosower:2003bh,GehrmannDeRidder:2005cm}.
A large effort has been made to investigate $\ttbar$ production at threshold, presently known at next-to-next-to-next-to-leading order QCD~\cite{Beneke:2015kwa}. 
For $b$ quarks, the next-to-next-to-leading order (NNLO) corrections (corresponding to $\mathcal{O}(\alpha^2_s)$ for this process) were calculated previously in the limit of vanishing $b$-quark mass~\cite{Ravindran:1998jw,Catani:1999nf,Weinzierl:2006yt}, and later with full $b$ quark mass dependence accounted for in ref.~\cite{Bernreuther:2016ccf}.

The talk is on the QCD corrections to the production of top and bottom quarks at $e^+e^-$ colliders in the continuum, based on the work reported in~refs.\cite{Chen:2016zbz,Bernreuther:2016ccf,Wang:2020ell}.
Namely, we analyze heavy $Q{\bar Q}$ production in $e^+e^-$ collisions, 
\begin{equation} \label{eq:QQincl}
 e^+ e^- \to \gamma^*,Z^* \to Q~{\bar Q} + X  \, ,
\end{equation}   
at the differential level to $\alpha_s^2$ in pertrubative QCD and apply it to $\ttbar$ production above the pair-production threshold and to $\bbbar$ production at the $Z$ resonance.

\section{Ingredients for computing $e^+e^- \rightarrow Q \bar{Q} ~+~ X$ at ${\cal O}(\as^2)$}
\label{SEC:Ingredients}

\subsection*{Matrix elements to ${\cal O}(\as^2)$}

To order $\as^2$, the (differential) cross section of the reaction~(\ref{eq:QQincl}) receives contributions from 
\begin{itemize} 
\item[i)] 
the two-parton $Q \bar Q$ state (at Born level, to order $\alpha_s$, and to order $\alpha_s^2$), 
\item[ii)] the three-parton state  $Q {\bar Q} g$  (to order $\alpha_s$ and to order $\alpha_s^2$),
\item[iii)] and the four-parton states $Q{\bar Q}gg$,  $Q{\bar Q} q \bar q$, and above the $4Q$-threshold from $Q{\bar Q}Q{\bar Q}$ 
(to order  $\alpha_s^2$).
\end{itemize} 
Contributions from the $Q{\bar Q}Q{\bar Q}$ final state should be treated with care, in view of the fact that for these states the  particle-multiplicity $n_Q \neq 1$ if one considers so-called inclusive heavy-quark distributions $d\sigma(e^+e^-\to Q + X)/dO_Q$, where $O_Q$ is some observable associated with $Q$.

\begin{figure}[htbp]
\begin{center}
\includegraphics[width=8.5cm,height=6.2cm]{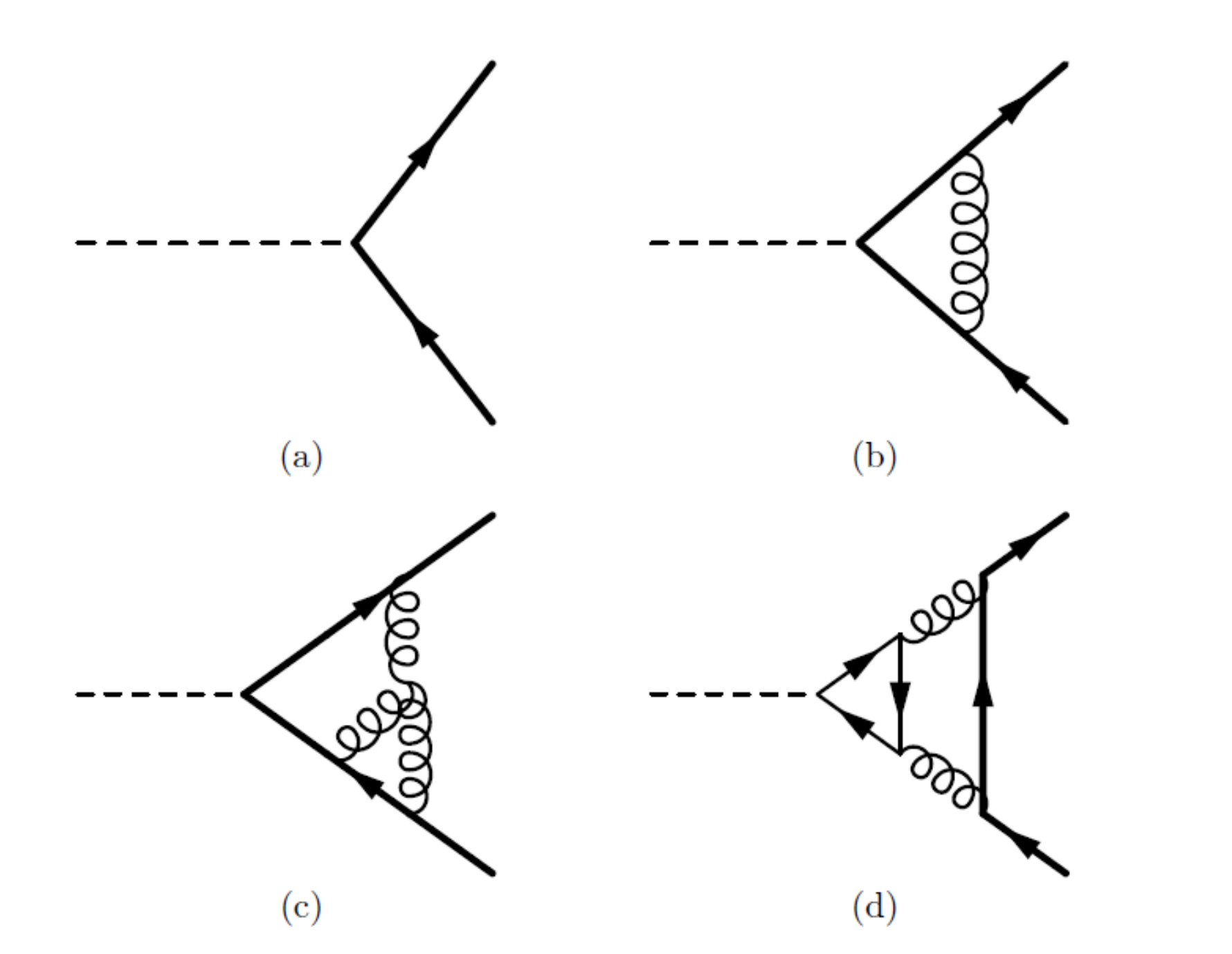}
\caption{Examples of diagrams that contribute to the $Q{\bar Q}$ 
final state to order $\as^2$. 
The dashed line represents the 
electroweak neutral current, that is, the virtual photon or $Z$ boson, the thick line the massive $Q$ quark, 
and the thin line any of the six quarks. 
The triangle diagrams (d) are summed over the six quark flavors.}
\label{fig:feynman1}
\end{center}
\end{figure}

We start with the two-parton $Q \bar Q$ contributions. 
Fig.~\ref{fig:feynman1} shows representative Feynman diagrams for this two-parton final state up to $\alpha_s^2$. 
According to the classification used in \cite{Bernreuther:2004ih,Bernreuther:2004th,Bernreuther:2005rw},
the two-loop diagram fig.~\ref{fig:feynman1}c belongs to the type-A two-loop contributions where the external current couples to 
to $Q{\bar Q}$. This diagram and the one-loop diagram fig.~\ref{fig:feynman1}c are also 
examples of so-called {\it universal QCD corrections} because the same electroweak couplings as in the lowest order diagram fig.~\ref{fig:feynman1}a are involved. 
Fig.~\ref{fig:feynman1}d is an example of the so-called type-B two-loop contributions where a fermion triangle loop is involved. 
Since in fig.~\ref{fig:feynman1}d it is not necessarily the heavy quark $Q$ that is coupled to the electroweak current, it is an example for the so-called {\it non-universal QCD corrections} to the lowest order $Q{\bar Q}$ production amplitude.

\begin{figure}[htbp]
\begin{center}
\includegraphics[width=12cm,height=4.3cm]{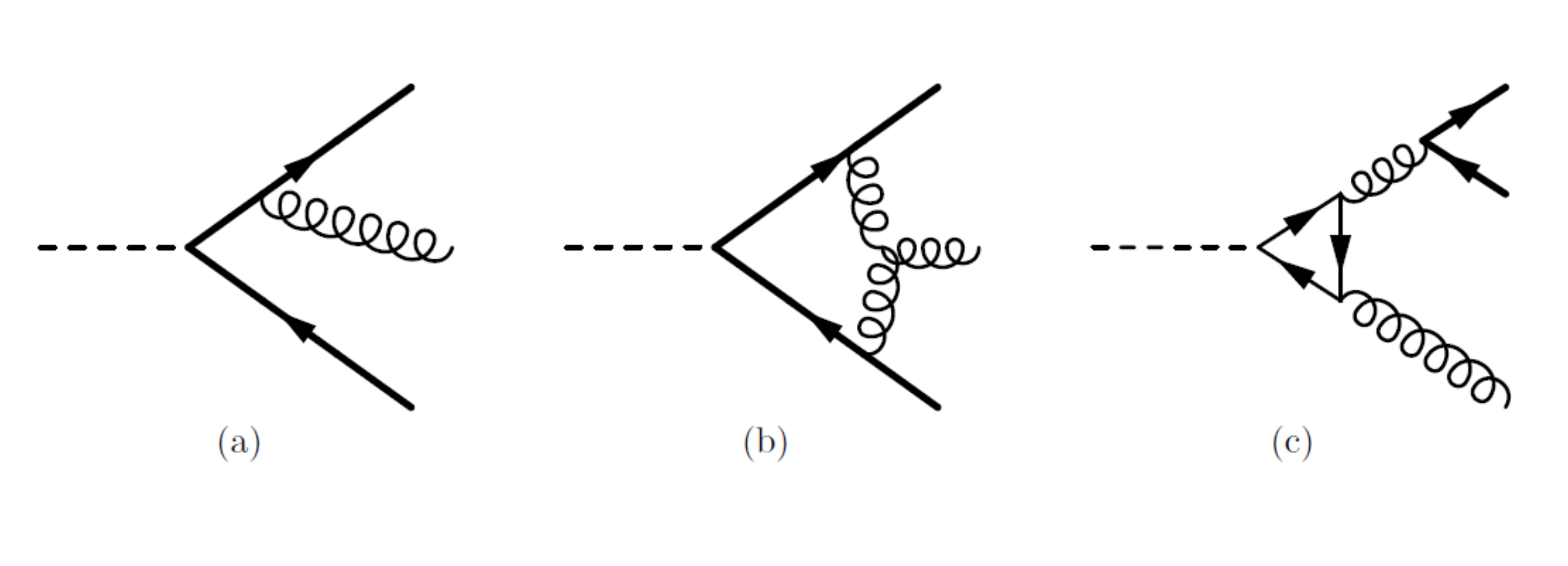}
\caption{Examples of diagrams that contribute to the 
$Q {\bar Q} g$ final state to order $\as^2$. 
The assignment of the lines is as in  fig.~\ref{fig:feynman1}.}
\label{fig:feynman2}
\end{center}
\end{figure}

\begin{figure}[htbp]
\begin{center}
\includegraphics[width=12cm,height=4.1cm]{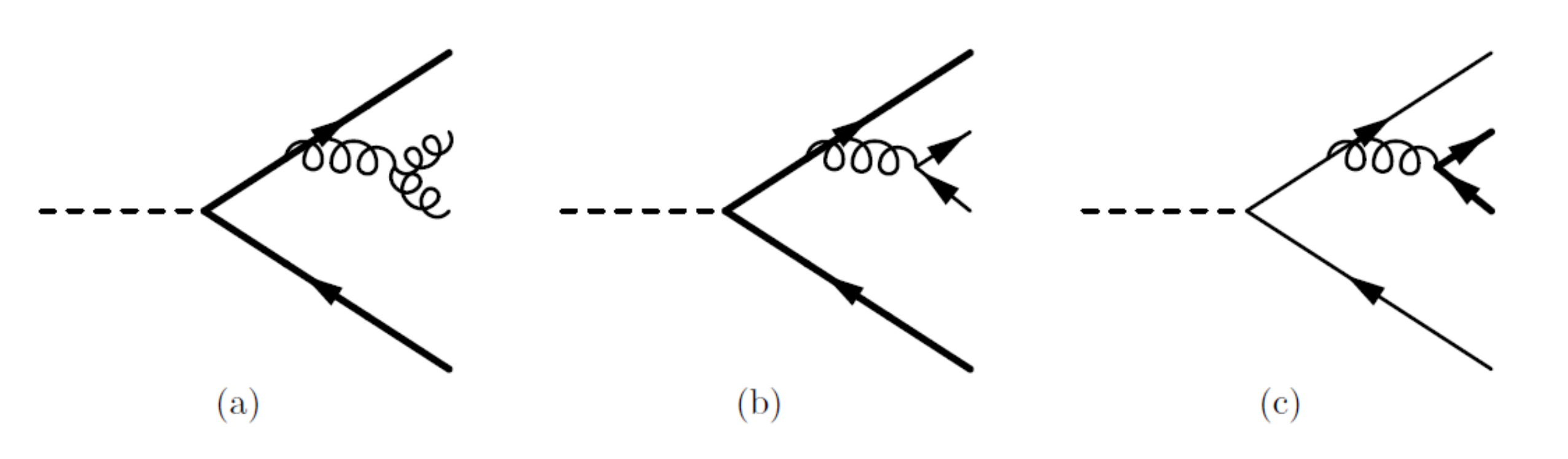}
\caption{(a): Examples of diagrams that contribute 
to the $Q{\bar Q}gg$ final state at order $\as^2$. 
(b,c): Two of the four diagrams that contribute to the $Q{\bar Q}q{\bar q}$ $(q\neq Q)$ 
final state at order $\as^2$.}
\label{fig:feynman3}
\end{center}
\end{figure}

Examples of diagrams associated with the $Q{\bar Q}g$ final state that lead to contributions of order $\as$ and $\as^2$ to the differential cross section of \eqref{eq:QQincl} are displayed in fig.~\ref{fig:feynman2}.
The diagrams fig.~\ref{fig:feynman2}a,b and fig.~\ref{fig:feynman2}c belong to the universal and non-universal QCD corrections, respectively.
The four-parton final-state diagrams of fig.~\ref{fig:feynman3} contribute at order $\as^2$ to the differential cross section.
An example of a diagram corresponding to the final state $Q{\bar Q}gg$ is shown in fig.~\ref{fig:feynman3}a,
while two of the four diagrams associated with  $Q{\bar Q}q{\bar q}$ $(q\neq Q)$ are exhibited in fig.~\ref{fig:feynman3}b,c.
The square of the  $Q{\bar Q}gg$ diagrams and the square of fig.~\ref{fig:feynman3}a lead to universal QCD corrections while the square of fig.~\ref{fig:feynman3}c and the interference of figs.~\ref{fig:feynman3}b and~\ref{fig:feynman3}c belong to the non-universal QCD corrections.
The calculations of these tree-level and one-loop Feynman diagrams are nowadays rather straightforward.

\subsection*{The antenna subtraction terms}

The antenna subtraction method is a systematic (process-independent) procedure for the construction of infrared subtraction terms that was developed for NLO QCD calculations in refs.~\cite{Kosower:1997zr,Kosower:2003bh} and has been generalized to NNLO QCD in ref.~\cite{GehrmannDeRidder:2005cm}. 
The antenna subtraction method is based on the use of color-ordered amplitudes into which an $n$-point parton amplitude at $l$ loops, $M_n^{(l)}$, can be decomposed\cite{Parke:1986gb,Mangano:1987xk,Berends:1987cv}:
$M_n^{(l)} = \sum_i C_{i,n}^{(l)} {\cal M}^{(l)}_i$ where $i$ runs over the $(n-1)!$ non-cyclic permutations of the external parton spin and momentum labels. 
This means that the ordering of the external partons is fixed in the color-stripped partial amplitudes ${\cal M}^{(l)}_i$. 
The coefficients $C_{i,n}^{(l)}$ contain the color information.
The development of the antenna subtraction method relies on two important features. 
First, the construction of subtraction terms for the color-ordered partial amplitudes is considerably simpler than for the full amplitudes. 
It relies on the IR factorization properties of the color-ordered partial amplitudes~\cite{Parke:1986gb,Mangano:1987xk,Berends:1988zn,Mangano:1990by,Campbell:1997hg,Kosower:1997zr,Bern:1999ry,Kosower:1999xi}).
The determination of the integrated subtraction terms in analytic fashion in $D\neq 4$ dimensions takes advantage of the modern multi-loop integration techniques. 
It has been accomplished at NNLO QCD for the massless case~\cite{Gehrmann-DeRidder:2003pne,GehrmannDeRidder:2005cm}, but also for a number of cases involving massive quarks~\cite{GehrmannDeRidder:2009fz,Abelof:2011jv,Abelof:2011ap,Abelof:2012he,Abelof:2014fza,Abelof:2014jna,Abelof:2015lna}.
For the computation of the reaction~(\ref{eq:QQincl}) at the differential level, we use the unintegrated and integrated NNLO real radiation antenna subtraction terms and the NNLO real-virtual antenna functions worked out in~\cite{Bernreuther:2011jt,Bernreuther:2013uma,Dekkers:2014hna}. 
For a synopsis of these objects needed, we refer to ref.~\cite{Chen:2016zbz}.

\section{Application to the top quark pair prodution in the continuum}
\label{SEC:ttbar}

\begin{figure}[tbh!]
\centering
\includegraphics[width=12cm,height=8.5cm]{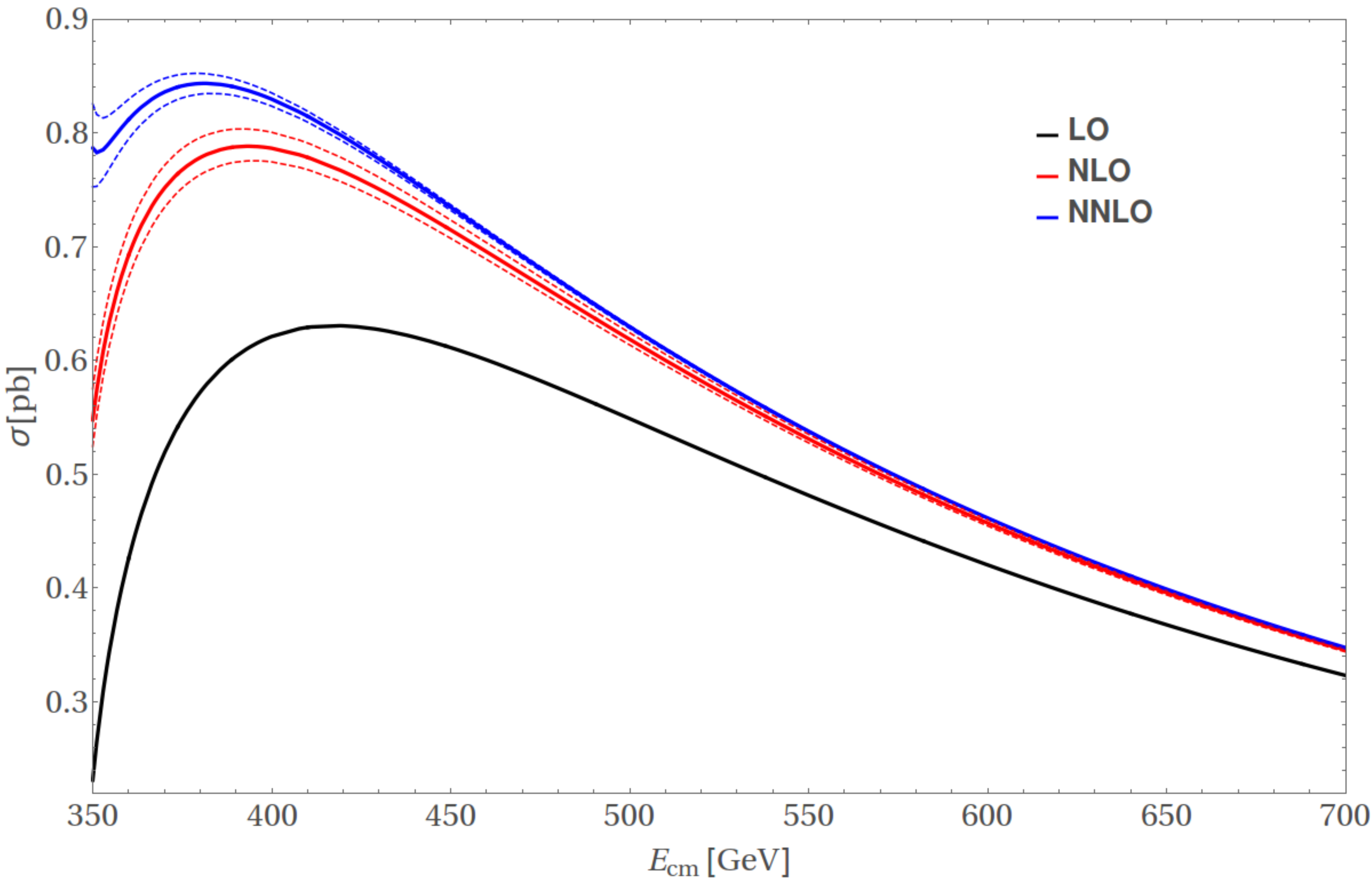}
\caption{The total $\ttbar$ cross section at LO, NLO, and NNLO QCD as a function of the c.m. energy.
The solid lines that refer to $\sigma_{\rm NLO}^{\ttbar}$ and $\sigma_{\rm NNLO}^{\ttbar}$
correspond to the choice $\mu=\sqrt{s}$, the dashed lines correspond to 
$\mu=\sqrt{s}/2$ and $2\sqrt{s}$.}
\label{fig:sigtt}
\end{figure}

We now apply our computational set-up to the study of the top quark pair production for unpolarized $e^+$ and $e^-$ beams and  $e^+e^-$ center-of-mass (c.m.) energies $\sqrt{s}$ sufficiently away from the $\ttbar$ threshold, where the fixed order perturbation theory in $\as$ is applicable. 
Fig.~\ref{fig:sigtt} 
shows our result for the $e^+e^- \rightarrow \ttbar$ cross section at LO, NLO, and NNLO QCD for c.m. energies $\sqrt{s} \lesssim 4 m_t$ (with input values of the SM parameters specified in ref.~\cite{Chen:2016zbz}).
In the case of $\sigma_{\rm NLO}^{\ttbar}$ and $\sigma_{\rm NNLO}^{\ttbar}$ the solid lines represent the values computed with the choice $\mu=\sqrt{s}$ for the renormalization scale. 
Uncertainties due to undetermined higher-order corrections are conventionally estimated by varying the renormalization-scale $\mu$ between $\sqrt{s}/2$ and $2\sqrt{s}$.
The upper and lower dashed lines correspond to these scale variations. 
In order to exhibit the size of the higher order QCD corrections relative to the LO cross section, we represent the $\ttbar$ cross section in the form
\begin{equation} \label{eq:sigDel12}
\sigma_{\rm NNLO}^{\ttbar} = \sigma_{\rm LO}^{\ttbar} \,(1 + \Delta_1 +  \Delta_2 ) \, .
\end{equation}
We list, for a reference, in table~\ref{tab:delta12}
\vspace{2mm}
\begin{table}[tbh!]
\begin{center}
\caption{\label{tab:delta12} The QCD corrections  
$\Delta_1$ and $\Delta_2$ to the LO ${\ttbar}$ cross section defined in Eq.~\eqref{eq:sigDel12} for 
several c.m. energies and  $\mu=\sqrt{s}$.}
\vspace{1mm}
\begin{tabular}{|c|cccc|}\hline
$\sqrt{s}$ [GeV] & 360 & 381.3 & 400  & 500 \\  \hline 
$\Delta_1$  & 0.627 & 0.352& 0.266 & 0.127 \\
$\Delta_2$ & 0.281 & 0.110 & 0.070 & 0.020 \\ \hline
\end{tabular}
\end{center}
\end{table}
the QCD corrections $\Delta_1$ and $\Delta_2$ for selected c.m. energies $\sqrt{s}$ for $\mu=\sqrt{s}$.

From figures~\ref{fig:sigtt} we can see that the higher order QCD corrections to the $\ttbar$ cross sections decrease with increasing c.m. energies $\sqrt{s}$ after $\sim 380$ GeV. 
As clearly shown, in the region where the perturbative calculation is reliable, the scale-variation band at NNLO is narrower than at NLO, as expected in perturbative calculations.
Notice also that the scale-variation bands do not overlap here. 
In the near threshold region, the QCD corrections become rather large. 
The blue thick line in fig.~\ref{fig:sigtt} representing the total NNLO result already shows a hint of blowing up when $\sqrt{s}\to 2 m_t.$ 
This is due to the well-known Coulomb singularities~\cite{Smith:1993vp,Chetyrkin:1996cf} in this process.
The NLO cross section, i.e.,~the red line in fig.~\ref{fig:sigtt}, does not diverge but goes to a constant for
$\beta \rightarrow 0$, because it involves only a single ${1}/{\beta}$ Coulomb singularity, due to one virtual longitudinal gluon exchange, which is compensated by the suppression factor $\beta$ due to the shrinking phase-space. 
The presence of Coulomb singularities in fixed order perturbative calculations near threshold clearly signifies the break-down of such a calculation. 
Within perturbation theory, one way to improve in the near threshold region is to sum up these Coulomb singularities through all perturbative orders, which then leads to a finite result (see, for instance,~\cite{Smith:1993vp,Brambilla:2004jw,Beneke:2015kwa}).

The numbers in table~\ref{tab:delta12} suggest that fixed order perturbation theory can be applied for $\sqrt{s} > 360$ GeV.
In the computation of $\sigma_{\rm NNLO}^{\ttbar}$, we have included also the non-universal contributions of order $\mathcal{O}(\as^2)$ that contain the electroweak couplings of quarks $q\neq t$. 
These contributions are free of divergences and are very small. 
For instance, at $\sqrt{s}$ = 500 GeV they amount to $-0.16\%$ of the total second order correction $\Delta_2$ defined in Eq.~\eqref{eq:sigDel12}, and this fraction decreases further in magnitude for smaller c.m. energies.

We have numerically compared our exact NNLO result with the expanded results in the near threshold region~\cite{Czarnecki:1997vz,Beneke:1997jm,Hoang:1997sj,Bernreuther:2006vp} and in the asymptotic region~\cite{Gorishnii:1986pz,Chetyrkin:1996cf,Chetyrkin:1997qi,Chetyrkin:1997pn} where $m_t^2/s\ll 1$, and found agreement. 
Furthermore, our results shown in table~\ref{tab:delta12} agree with the calculation of the $\ttbar$ cross section in~ref.\cite{Gao:2014eea}. 
For a detailed discussion of our results on differential distributions of the top quark pair prodution we refer to ref.~\cite{Chen:2016zbz}.

\section{The forward-backward asymmetries in $e^+ e^- \to Q {\bar Q}$}
\label{SEC:AFB}

\subsection{Definitions of the un-expanded and expanded $A_{\rm FB}$}

The forward-backward asymmetry $A_{\rm FB}$ in the heavy $Q{\bar Q}$ production, is defined by
\begin{eqnarray}
A_{\rm FB} \equiv \frac{N_F - N_B}{N_F + N_B} \, ,
\label{AFBdef}
\end{eqnarray}
where $N_F$ $(N_B)$ is the number of quarks $Q$ produced in the forward (backward) direction. 
The asymmetry $A_{\rm FB}$ in the reaction $e^+ e^- \to Q {\bar Q} + X$ is generated by those terms in the squared S-matrix elements which are odd under the interchange of $Q$ and $\bar Q$ while the initial state is kept fixed. 
The identification of the forward/backward direction involves a choice of reference axis. 
The definition of the reference axis must be such that the resulting forward-backward asymmetry is an infrared safe (IR-safe) quantity so that it can be reliably calculated and subsequently compared with experimental measurements. 
Common choices include the direction of flight of the heavy quark and the oriented thrust axis defined by a certain thrust-finding algorithm~\cite{Farhi:1977sg,Brandt:1964sa}.

The heavy quark forward-backward asymmetry $A_{\rm FB}$ can also be expressed in terms of the symmetric cross section $\sigma_S$ and the antisymmetric cross section $\sigma_A$ for the inclusive production of the heavy quark $Q$, i.e.,
\begin{eqnarray} \label{defafb}
A_{\rm FB}=\frac{\sigma_A}{\sigma_S}=\frac{\sigma_F-\sigma_B}{\sigma_F+\sigma_B}.
\end{eqnarray}
The $\sigma_F$ and $\sigma_B$ are the forward and backward cross sections, respectively, which can be written in terms of differential cross sections as
\begin{eqnarray}
\sigma_F=\int_0^1d\cos\theta \int_{x_0}^1dx\frac{d\sigma}{dx\,d\cos\theta}\,, \,\,
\sigma_B=\int_{-1}^0d\cos\theta \int_{x_0}^1dx\frac{d\sigma}{dx\,d\cos\theta},
\end{eqnarray}
where $\theta$ is the angle between the electron three-momentum and the axis defining the forward hemisphere. 
The energy ratio $x$ is defined as $2\,E_{Q}/\sqrt{s}$ where $E_{Q}$ is the energy of the heavy quark $Q$.

To order $\alpha_s^2$ the symmetric and antisymmetric cross sections receive the following perturbative contributions:
\begin{align}\label{sasbeitraege}
\sigma_{A,S} = \sigma_{A,S}^{(2,0)} +\sigma_{A,S}^{(2,1)} + \sigma_{A,S}^{(3,1)} + \sigma_{A,S}^{(2,2)} 
+ \sigma_{A,S}^{(3,2)} + \sigma_{A,S}^{(4,2)} + {\cal O}(\as^3) \, ,
\end{align}
where the first number in the superscripts $(i,j)$  denotes the number of final-state partons associated with the respective  term and the second one the order of $\alpha_s$. 
Inserting (\ref{sasbeitraege}) into (\ref{defafb}) we get the unexpanded $A^b_{\rm FB}$ to first and to second order in $\as$:
\begin{align}
\afb(\as) = & \frac{\sigma_{A}^{(2,0)} +\sigma_{A}^{(2,1)}  + \sigma_{A}^{(3,1)}}{{\sigma_{S}^{(2,0)} 
 +\sigma_{S}^{(2,1)} + \sigma_{S}^{(3,1)}}} \equiv  \afb^{\rm LO}~C_1\, , \label{afb1unex} \\
\afb(\as^2) = & \frac{\sigma_{A}^{(2,0)} +\sigma_{A}^{(2,1)}  + \sigma_{A}^{(3,1)} + \sigma_{A}^{(2,2)}+ \sigma_{A}^{(3,2)} +
\sigma_{A}^{(4,2)}}{\sigma_{S}^{(2,0)} +\sigma_{S}^{(2,1)} + \sigma_{S}^{(3,1)}  + \sigma_{S}^{(2,2)} + \sigma_{S}^{(3,2)} +
\sigma_{S}^{(4,2)}} \equiv  \afb^{\rm LO}~C_2\, , \label{afbgesamt}
\end{align}
where 
\begin{equation}
 \afb^{\rm LO} = \frac{\sia^{(2,0)}}{\sis^{(2,0)}} \, ,
\label{eq:afb0}
\end{equation}
is the forward-backward asymmetry at Born level. 
The factors $C_1$ and $C_2$ defined by the respective ratio on the left of eq.~\eqref{afb1unex} and~\eqref{afbgesamt} are the unexpanded first-order and second-order QCD correction factors.

Taylor expanding eq.~\eqref{afb1unex} to first order and of~\eqref{afbgesamt} to second order in  $\as$ gives 
\begin{align}
\afb^{\rm NLO}= & \afb^{\rm LO} \left[1+A_1\right] \;+\; {\cal O}(\as^2) \, , \label{afb1ent} \\
\afb^{\rm NNLO} =&
\afb^{\rm LO}\;\left[1\;+\;A_{1} \;+\;
A_{2}\right]   \;+\; {\cal O}(\as^3)     \, , \label{afbent}
\end{align}
where $A_{1}$ and $A_{2}$ are the QCD corrections of ${\cal O}(\as)$ and  ${\cal O}(\as^2)$, respectively.
\begin{align}
A_{1} =& \sum\limits_{i=2,3} \big[\frac{\sia^{(i,1)}}{\sia^{(2,0)}} \;-\;\frac{\sis^{(i,1)}}{\sis^{(2,0)}} \big] \, ,\label{eq: afb1} \\
A_{2} =& \sum\limits_{i=2,3,4} \big[ \frac{\sia^{(i,2)}}{\sia^{(2,0)}} \;-\;
\frac{\sis^{(i,2)}}{\sis^{(2,0)}} \big]
 - \frac{\sis^{(2,1)}+\sis^{(3,1)}}{\sis^{(2,0)}} \, A_1  \, .\label{eq:afb2}
\end{align}
Eqs.~\eqref{afb1ent} and~\eqref{afbent} are the expanded forms of the forward-backward asymmetry at NLO and NNLO QCD.
The unexpanded and expanded first and second-order forward-backward asymmetries differ by terms of order $\as^2$ and order $\as^3$, respectively. 
The differences between the two forms may be considered as an estimate of the theory uncertainties.

\subsection{Numerical results for the $A_{\rm FB}^{t}$ in the continuum}

Table~\ref{tab:afbex} contains our results for the expanded form of the top-quark forward-backward asymmetry $A_{\rm FB}^{t}$ at NLO and NNLO QCD defined in eq.~\eqref{afb1ent} and~\eqref{afbent}, respectively, and the correction terms $A_1$ and $A_2$ given in Eq.~\eqref{eq: afb1} and~\eqref{eq:afb2} for several c.m. energies. 
The central values and the uncertainties refer to the scales $\mu=\sqrt{s}$ and $\mu=\sqrt{s}/2$ and $2\sqrt{s}$, respectively.
Obviously, here we vary $\mu$ simultaneously in the $\sigma_S$ and $\sigma_A$ of different partonic channels between $\sqrt{s}/2$ and $2\sqrt{s}$.
We note that  we have included in $A_2$ also the non-universal contributions that contain the electroweak couplings of quarks $q\neq t$.
\vspace{2mm}
\begin{table}[tbh!]
\begin{center}
\caption{\label{tab:afbex} The top-quark forward-backward asymmetry at LO, NLO, and NNLO QCD for several c.m. energies using the expansions in eq.~\eqref{afb1ent} and~\eqref{afbent}. The numbers are given in percent.}
\vspace{1mm}
\begin{tabular}{|c|ccc|cc|c|}\hline
  $\sqrt{s}$ [GeV] & $\afb^{\rm LO} $ [\%]  & $\afb^{\rm NLO}$ [\%] & $\afb^{\rm NNLO}$ [\%] & $A_1$ [\%] & $A_2$  [\%] & $\delta\afb^{\rm NNLO}$ [\%]\\  \hline 
    360 & 14.94 & $15.54^{+0.05}_{-0.04}$ & $16.23^{+0.12}_{-0.10}$ & $4.01^{+0.35}_{-0.29}$ &  $4.58^{+0.46}_{-0.38}$ & $\pm  0.59  $ \\[.3em]
    400 & 28.02 & $28.97^{+0.08}_{-0.07}$ & $29.63^{+0.11}_{-0.10}$ & $3.41^{+0.29}_{-0.25}$ &  $2.36^{+0.11}_{-0.11}$ & $\pm  0.27 $ \\[.3em]
    500 & 41.48 & $42.42^{+0.08}_{-0.07}$ & $42.91^{+0.08}_{-0.07}$ & $2.28^{+0.19}_{-0.16}$ &  $1.18^{+0.01}_{-0.01}$ & $\pm  0.13 $ \\[.3em] 
    700 & 51.34 & $51.81^{+0.04}_{-0.03}$ & $52.05^{+0.04}_{-0.04}$ & $0.91^{+0.07}_{-0.06}$ &  $0.47^{+0.01}_{-0.01}$ & $\pm  0.06 $ \\[.3em] \hline
\end{tabular}
\end{center}
\end{table}
Table~\ref{tab:afbunex} contains our results for the unexpanded form of the top-quark $A^{t}_{\rm FB}$ at NLO and NNLO QCD and the associated QCD correction factors $C_1$ and $C_2$ defined in eq.~\eqref{afb1unex} and~\eqref{afbgesamt} for several c.m. energies. 
This form of computing the asymmetry corresponds to the simulation using Monte-Carlo event generators.
The $A^{t}_{\rm FB}$ at NNLO QCD was computed before in ref.~\cite{Gao:2014eea} in the unexpanded form with values of $m_t$ and $\alpha_s$ that differ slightly from the ones that we use here, while the results basically agree.
\vspace{2mm}
\begin{table}[tbh!]
\begin{center}
\caption{\label{tab:afbunex} The unexpanded form of the top-quark forward-backward asymmetry \eqref{defafb} for several c.m. energies.
 The numbers are given in percent.}
\vspace{1mm}
\begin{tabular}{|c|ccc|}\hline
  $\sqrt{s}$ [GeV] & $\afb^{\rm LO} $  [\%]     & $\afb^{\rm NLO}$ [\%] & $\afb^{\rm NNLO}$ [\%] \\  \hline 
    360            & 14.94 & $15.31^{+0.02}_{-0.02}$  & $15.82^{+0.08}_{-0.06}$  \\[.3em]
    400            & 28.02 & $28.77^{+0.05}_{-0.04}$  & $29.42^{+0.10}_{-0.09}$  \\[.3em]
    500            & 41.48 & $42.32^{+0.06}_{-0.05}$  & $42.83^{+0.08}_{-0.07}$  \\[.3em] 
    700            &  51.34 &  $51.78^{+0.03}_{-0.03}$ & $52.03^{+0.04}_{-0.04}$ \\[.3em] \hline          
\end{tabular}
\end{center}
\end{table}
One may take the spread between the values of the expanded and unexpanded forms of $A_{FB}^{\rm NLO}$  and  $A_{FB}^{\rm NNLO}$ given in tables~\ref{tab:afbex},~\ref{tab:afbunex} as an estimate of the uncalculated higher order corrections.

\subsection{Numerical results for the $A_{\rm FB}^{b}$ at the $Z$ resonance}

We now present our numerical results on the $b$-quark forward-backward asymmetry at the $Z$ resonance to $\as^2$ in QCD and lowest order in the electroweak couplings.
The input values of the SM couplings and masses are specified in detail in ref.~\cite{Bernreuther:2016ccf}.
In particular, we use $\as$ defined in the 5-flavor QCD and the bottom-quark on-shell mass $m_b= 4.89$~GeV.
For the mass of the top quark that appears in the triangle-loop diagrams contributing to the non-universal corrections to $A_{\rm FB}^{b}$ we use $m_t=173.34$ GeV. The $u,d,c,s$ quarks are taken to be massless.
With these input parameters, the Born-level value of the $b$-quark forward-backward asymmetry at $\sqrt{s}=m_Z$  is $\afb^{\rm LO}=0.1512$. 
The value of $\afb^{\rm LO}$ is very sensitive to the input value of $s^2_W$ but insensitive to the value of $m_b$ within its uncertainty.
Here, we discuss only the results of the QCD corrections for the expanded $A_{\rm FB}^{b}$ with respect to both the 
$b$-quark and the thrust direction, the later of which enters the subsequent discussion of the phenomenological consequences.
Table~\ref{tab:afbb-exp}
lists the values of the NLO and NNLO correction factor, both for the quark axis and the thrust axis definition, for the three 
renormalization scales $\mu=m_Z/2, m_Z, 2 m_Z$.
The numbers given in this table show that the order $\as^2$ corrections are significant. 
For $\mu=m_Z$ the ratio $A_2/A_1$ is $43\%$ and $37\%$ for the quark and thrust axis definition, respectively. 
The scale variations change both the first-  and second-order QCD correction factors by about $\pm 0.003$ with respect to their values at $\mu=m_Z$.
The fact that inclusion of the second-order correction term $A_2$ does not reduce the scale uncertainty is not unusual for an observable that is defined as a ratio.
Contrary to the case of the top quark for the c.m. energies below the $4 m_t$ threshold, the order $\alpha_s$ and $\alpha^2_s$ corrections $A_1$ and $A_2$ are dominated by the contributions from the three-parton and three- and four-parton final states, 
respectively.
We have included in the computation of $\afb$ the non-universal corrections of order $\as^2$ that contain the vector and axial vector couplings of quarks $q\neq b$, which turn out to be quite sizable. 
We note that the non-universal corrections depend on the value of $\sin^2\theta_W$.
\begin{table}[tbh!]
\begin{center}
\caption{\label{tab:afbb-exp} The first- and second-order QCD correction factors defined in \eqref{afb1ent} - \eqref{eq:afb2} 
to the LO $b$-quark forward-backward asymmetry at the $Z$ peak $\mu=m_Z$. 
The numbers in superscript (subscript) refer to the changes if $\mu=2 m_Z$  ($\mu=m_Z/2$) is chosen.}

\vspace{1mm}
 \begin{tabular}{| c|c|c|c|c|}\hline 
 &     $1+A_1$& $1+A_1+ A_2$  & $A_1$ & $A_2$ \\ \hline 
 quark axis: & $0.9710^{+0.0028}_{-0.0034}$ & $0.9587^{+0.0026}_{-0.0028}$ & $ -0.0290$ &  $-0.0123$ \\ [2mm] \hline                             
  thrust axis: &  $0.9713^{+0.0027}_{-0.0026}$ & $0.9608^{+0.0022}_{-0.0025}$  & $-0.0287$ & $-0.0105$ \\[2mm] \hline    
 \end{tabular}
 \end{center}
 \end{table}

\subsection{Phenomenological consequences}

The $A_{\rm FB}$ with QCD corrections calculated above cannot be compared directly with experimental measurements. 
In the measurements of the $b$-quark asymmetry by the experiments at the previous $e^+e^-$ colliders LEP and SLAC, which are reviewed in~\cite{ALEPH:2005ab,Group:2009ae,ALEPH:2010aa}, the forward and backward hemispheres was defined using the thrust axis.  
In our computation the thrust axis is used but identified only for partonic final states, while the hadronization of partons causes a smearing of this axis (causing discrepancy between thrust at the parton level and at the level of stable hadrons). 
Results from several experimental measurements with different acceptance regions in phase space, and different sensitivity to QCD-correction effects etc, are combined after unfolded respectively.
Bearing this in mind, we proceed with the analysis as described in ref.~\cite{Bernreuther:2016ccf}, which, albeit being very preliminary and far from full-fledged, can give a quick idea about what might be the possible consequence of a revised QCD-correction factor~\footnote{See ref.~\cite{dEnterria:2020cgt} for an updated analysis of the theoretical uncertainties.}. 
In short, a pseudo-observable, the bare $b$-quark asymmetry $A^{0,b}_{\rm FB}$, is extracted from the measured asymmetry $A^{b,T}_{\rm FB,exp}$~\cite{LEPHF,Abbaneo:1998xt} in the following way.  
The the measured asymmetry $A^{b,T}_{\rm FB,exp}$ was first corrected for QCD effects as follows, 
\begin{eqnarray}
A^{b,T}_{\rm FB,exp}&=&(A^{0,b}_{\rm FB})_{\rm exp}\left[1+\delta^{(1)}_Ta_s+\delta^{(2)}_Ta_s^2\right]\,,
\label{EQ:AFBexp}
\end{eqnarray}
and then the QCD corrected ``experimental'' asymmetry $(A^{0,b}_{\rm FB})_{\rm exp}$ was further corrected for 
higher order electroweak corrections etc, before a value of the bare asymmetry $A^{0,b}_{\rm FB}$ was deduced: 
\begin{equation} \label{eq:finalcorr}
 A_{\rm FB}^{0,b} = (\afb^{0,b})_{\rm exp} + \delta \afb^b \, .
\end{equation}
The corrections $\delta \afb^b$ include the energy shift from $\sqrt{s}=91.26$ GeV (all measured asymmetries were corrected, in a first step, to this energy) to $\sqrt{s}= m_Z$, QED corrections, corrections due to $\gamma$ exchange and $Z-\gamma$ interference, and due to imaginary parts of effective weak couplings (cf. for instance,~\cite{Freitas:2004mn}). 
In this way, the experimental value determined for this pseudo-observable reads as $A^{0,b}_{\rm FB}=0.0992\pm0.0016$ and shows a large deviation, about $2.9\sigma$, from the SM fit $A^{0,b}_{\rm FB}=0.1038$~\cite{ALEPH:2005ab,ALEPH:2010aa} if the NNLO QCD corrections for massless $b$ quarks are used.

The QCD correction factor determined in~\cite{LEPHF} is $(1-C^T_{\rm QCD})=0.9646 \pm 0.0063$ where the error includes estimates of hadronization effects. 
Our thrust axis correction factor $(1+A_1+ A_2) = 0.9608 \pm 0.0025$ where the error is due to scale uncertainties only, agrees with that factor within the uncertainties. 
Our central value is smaller than $0.9646$ by $0.4\%$, and accordingly our correction changes the value of the pseudo-observable $A_{FB}^{0,b}$ to $0.0996 \pm 0.0016$.
Thus the pull between $A_{FB}^{0,b}$ and the SM fit cited above is slightly decreased, namely from $2.9\sigma$ to $2.6 \sigma$.

The fixed-order results discussed above were determined at the conventional scale choice $\mu = m_z$. 
Due to the missing higher order perturbative corrections, the truncated fixed order results on $ A_{\rm FB}^{b}$ still have a remaining $\mu$ dependence.
One way to improve the perturbative result regarding this aspect is to resum all explicit $\ln \mu^2$ terms in $ A_{\rm FB}^{b}$ that are related the QCD $\beta = \mathrm{d} \ln \alpha_s / \mathrm{d} \ln  \mu^2$ function, absorbed into a new effective coupling for this particular observable, by means of the well-known optimization procedure based on the Principle of Maximum Conformality (PMC). 
Following the PMC procedure we determined the PMC scale to be $\mu^{\rm PMC}_r =9.7 ~\rm {GeV}$ for the $b$ quark forward-backward asymmetry with the thrust axis definition in ref.~\cite{Wang:2020ell}. 
After applying PMC scale setting, the QCD correction factor with the thrust axis definition is $(1+\delta^{(1)}_Ta_s+\delta^{(2)}_Ta_s^2)=0.9529$, which is smaller than $0.9646$ by $1.2\%$. 
Consequently, the QCD correction factor with refinements from the PMC method changes $A^{0,b}_{\rm FB}=0.0992\pm0.0016$ to
\begin{eqnarray}
A^{0,b}_{\rm FB}=0.1004\pm0.0016,
\end{eqnarray}
which shows a $2.1\sigma$ deviation from the SM fit $A^{0,b}_{\rm FB}=0.1038$.

It is noted that there are updated SM fit results reported in refs.~\cite{Baak:2014ora,Tanabashi:2018oca} with improved electroweak corrections taken into account. 
To see more clearly how the deviations are changed due to the incorporation of updated electroweak and QCD corrections, in  table~(\ref{tab:AFBnumbers}), we list the deviations computed using different central values of the SM fit $A^{b,\text{fit}}_{FB}$, and also the $A^{0, b}_{FB}$ determined by making use of the different parton-level QCD correction factors. 
\begin{table}
\begin{center}
\begin{tabular}{c||c c c} 
$A^{b,\text{fit}}_{FB}$ v.s. $A^{0, b}_{FB}$ & 0.0992~ref.~\cite{ALEPH:2010aa} & 0.0996~ref.~\cite{Bernreuther:2016ccf} & 1.004~ref.~\cite{Wang:2020ell} \\[.2em]
\hline
\hline
\\[.2em]
0.1038~ref.~\cite{ALEPH:2005ab} & 2.9 $\sigma$ & 2.6 $\sigma$ & 2.1 $\sigma$ \\[.2em]
\\[.2em]
0.1032~ref.~\cite{Baak:2014ora} & 2.5 $\sigma$ & 2.25 $\sigma$ & 1.8 $\sigma$ \\[.2em]
\\[.2em]
0.1030~ref.~\cite{Tanabashi:2018oca} & 2.4 $\sigma$ & 2.1 $\sigma$ & 1.6 $\sigma$ \\[.2em]
 \hline 
\end{tabular}
\caption{\label{tab:AFBnumbers} Deviations between different central values of the SM fit $A^{b,\text{fit}}_{FB}$ and the $A^{0, b}_{FB}$ determined in literature}
\end{center}
\end{table}
So far the tension between $A^{b,\text{fit}}_{FB}$ and $A^{0, b}_{FB}$ seems to decrease progressively over the years.
However, the experimental uncertainty of $ A_{\rm FB}^{b}$ measurement used in determining the numbers in this table is still the one at the SLAC and LEP experiments, about $1.7\%$~\cite{ALEPH:2005ab}. 
On the other hand, the experimental uncertainty, currently dominated by the statistical uncertainties, is expected to be greatly reduced at future lepton colliders~\cite{Hawkings:1999ac,Erler:2000jg,Devetak:2010na,Amjad:2013tlv,ILD:2020qve}.
Confirmation or resolution of this long-term discrepancy at the future lepton colliders would, of course, make a great impact on the development of our theories.

\section{Conclusions}
\label{SEC:conclusion}

We have formulated a set-up for computing the production of a heavy quark-antiquark pair $Q{\bar Q}$ in $e^+e^-$ collisions at $\mathcal{O}(\alpha_s^2)$ in the perturbative QCD at the differential level, using the antenna subtraction method for handling the intermediate infrared soft and collinear divergences of the individual contributions. 
This set-up applies to the differential calculation of any infrared-safe observable of this process to $\mathcal{O}(\alpha_s^2)$ in QCD, and works in a numerically stable way owing to the local subtraction of IR divergences in each partonic phase space.

Besides the application to $t\bar{t}$ pair production at (future) $e^+e^-$ collisions, we have determined the $\mathcal{O}(\alpha_s^2)$ QCD corrections to the b-quark forward-backward asymmetry $A_{\rm{FB}}$ in $e^+e^-$ collisions at the $Z$ resonance, aimed at analysing the long-standing tension between its direct determination at the SLAC and LEP experiments and its global SM fit.
We found that after taking into account exact b-quark mass effects, the pull between the direct determination of $A_{\rm{FB}}$ and its SM fit is reduced from 2.9$\sigma$ to 2.6$\sigma$. 
Furthermore, we performed an improved determination of this QCD correction factor by means of the optimization procedure based on the Principle of Maximum Conformality, with which the tension is eventually reduced down to 2.1$\sigma$.
Once compared to the updated SM-fit results, the deviation gets reduced below 2$\sigma$. 
As a future application of our computational set-up, one may consider the determination of the $\mathcal{O}(\alpha_s^2)$ corrections to the forward-backward asymmetry for two $b$-jet final states, for which one expects a decrease of the magnitude of the QCD corrections. 
Furthermore, this setup can be applied to the prodution of polarized top quarks, with polarized $\mathit{e}^+ \mathit{e}^-$ beams. 

\section*{Acknowledgments}

The author is grateful to Werner Bernreuther, Oliver Dekkers, Thomas Gehrmann, Dennis Heisler, Rui-qing Meng, Jian-ming Shen, Zong-guo Si, Sheng-quan Wang, Xing-gang Wu for collaborating on the work presented in this proceeding.

\bibliography{LCSW21_AFB} 
\bibliographystyle{utphysM}


\end{document}